\documentclass[english,aps,prl,reprint,superscriptaddress,longbibliography]{revtex4-2}

\usepackage[T1]{fontenc}
\usepackage[latin9]{inputenc}
\usepackage{amssymb}
\usepackage{graphicx}
\usepackage{subscript}
\usepackage{amsmath}
\usepackage[colorlinks=true,citecolor=blue,urlcolor=blue,linkcolor=blue]{hyperref}

\makeatletter
\usepackage{pslatex}

\usepackage{hyperref}% add hypertext capabilities
\usepackage{xcolor}
\hypersetup{
    colorlinks,
    linkcolor={blue!70!black},
    citecolor={blue!70!black},
    urlcolor={blue!70!black}
}

\usepackage{babel}
\usepackage{comment}

\newcommand{\X}{\tilde{X}^{2}\Sigma^{+}}
\newcommand{\A}{\tilde{A}^{2}\Pi_{1/2}}
\newcommand{\B}{\tilde{B}^{2}\Sigma^{+}}
\newcommand{\wn}{~\rm{cm}^{-1}}

\newcommand*{\At}{\ensuremath{\tilde{A}}}

\newcommand{\celltemp}{2.4} % Kelvin, checked
\newcommand{\peakv}{110} % m/s, checked
\newcommand{\widthv}{20} % m/s, checked using 240226 data
\newcommand{\maincycle}{23} % unitless, checked
\newcommand{\eomfreq}{4.4} % MHz, checked
\newcommand{\slowswitch}{1.1} % MHz, checked
\newcommand{\switchfreq}{1.4} % MHz, checked
\newcommand{\bgrad}{16} % G/cm (RMS amplitude), checked
\newcommand{\beamline}{145}
\newcommand{\enhanceP}{1}

\newcommand{\PquarterV}{1.1} % mW, checked
\newcommand{\PhalfV}{2.2} % mW, checked
\newcommand{\PoneV}{4.7} % mW, checked
\newcommand{\PtwoV}{9.4} % mW, checked

\newcommand{\MOTnumber}{2000}
\newcommand{\MOTnumberUnc}{600}

\newcommand{\tauAsymptote}{20}
\newcommand{\tauAsymptoteUnc}{5}
\newcommand{\Psaturate}{5}
\newcommand{\PsaturateUnc}{2}

\newcommand{\tempQuarterV}{1.2} %mK
\newcommand{\tempQuarterVUnc}{3}
\newcommand{\tempHalfV}{1.8}
\newcommand{\tempHalfVUnc}{3}
\newcommand{\tempOneV}{3.6}
\newcommand{\tempOneVUnc}{6}

\newcommand{\minCloud}{1.12} %mm
\newcommand{\minCloudUnc}{1}

\newcommand{\tauTwoV}{23}
\newcommand{\tauTwoVUnc}{1}
\newcommand{\tauOneV}{31}
\newcommand{\tauOneVUnc}{2}
\newcommand{\tauHalfV}{48}
\newcommand{\tauHalfVUnc}{4}
\newcommand{\tauQuarterV}{91}
\newcommand{\tauQuarterVUnc}{9}

\newcommand{\tauBFour}{44}
\newcommand{\tauBFourUnc}{3}
\newcommand{\tauBThree}{27}
\newcommand{\tauBThreeUnc}{3}
\newcommand{\tauBTwo}{23}
\newcommand{\tauBTwoUnc}{3}
\newcommand{\tauBOne}{12}
\newcommand{\tauBOneUnc}{2}

\newcommand{\tauRef}{\tauBFour}
\newcommand{\tauRefUnc}{\tauBFourUnc}
\newcommand{\tauShortCycle}{\tauBOne}
\newcommand{\tauShortCycleUnc}{\tauBOneUnc}

\newcommand{\cycle}{12600} % unitless, checked
\newcommand{\BOne}{3400}
\newcommand{\BTwo}{4300}
\newcommand{\BThree}{6100}
\newcommand{\BFour}{\cycle}

\newcommand{\betaHalfV}{113}
\newcommand{\betaHalfVUnc}{32}
\newcommand{\betaOneV}{112}
\newcommand{\betaOneVUnc}{32}
\newcommand{\betaTwoV}{55}
\newcommand{\betaTwoVUnc}{28}

\newcommand{\oscHalfV}{40}
\newcommand{\oscHalfVUnc}{2}
\newcommand{\oscOneV}{45}
\newcommand{\oscOneVUnc}{2}
\newcommand{\oscTwoV}{50}
\newcommand{\oscTwoVUnc}{2}

\newcommand{\numPulse}{$10^{10}$}

\newcommand{\photonsPerMolecule}{7}

\makeatother

\begin{document}

\title{Magneto-optical trapping of a heavy polyatomic molecule for precision measurement}

\author{Zack D. Lasner}
\affiliation{Harvard-MIT Center for Ultracold Atoms, Cambridge, Massachusetts 02138, USA}
\affiliation{Department of Physics, Harvard University, Cambridge, Massachusetts 02138, USA}

\author{Alexander Frenett}
\affiliation{Harvard-MIT Center for Ultracold Atoms, Cambridge, Massachusetts 02138, USA}
\affiliation{Department of Physics, Harvard University, Cambridge, Massachusetts 02138, USA}

\author{Hiromitsu Sawaoka}
\affiliation{Harvard-MIT Center for Ultracold Atoms, Cambridge, Massachusetts 02138, USA}
\affiliation{Department of Physics, Harvard University, Cambridge, Massachusetts 02138, USA}

\author{Lo\"{i}c Anderegg}
\altaffiliation{Current address: Department of Physics and Astronomy, University of Southern California, Los Angeles, California 90089, USA}
\affiliation{Harvard-MIT Center for Ultracold Atoms, Cambridge, Massachusetts 02138, USA}
\affiliation{Department of Physics, Harvard University, Cambridge, Massachusetts 02138, USA}

\author{Benjamin Augenbraun}
\altaffiliation{Current address: Department of Chemistry, Williams College, Williamstown, Massachusetts 01267, USA}
\affiliation{Harvard-MIT Center for Ultracold Atoms, Cambridge, Massachusetts 02138, USA}
\affiliation{Department of Physics, Harvard University, Cambridge, Massachusetts 02138, USA}

\author{Hana Lampson}
\affiliation{Harvard-MIT Center for Ultracold Atoms, Cambridge, Massachusetts 02138, USA}
\affiliation{Department of Physics, Harvard University, Cambridge, Massachusetts 02138, USA}

\author{Mingda Li}
\affiliation{Harvard-MIT Center for Ultracold Atoms, Cambridge, Massachusetts 02138, USA}
\affiliation{Department of Physics, Harvard University, Cambridge, Massachusetts 02138, USA}

\author{Annika Lunstad}
\affiliation{Harvard-MIT Center for Ultracold Atoms, Cambridge, Massachusetts 02138, USA}
\affiliation{Department of Physics, Harvard University, Cambridge, Massachusetts 02138, USA}

\author{Jack Mango}
\affiliation{Harvard-MIT Center for Ultracold Atoms, Cambridge, Massachusetts 02138, USA}
\affiliation{Department of Physics, Harvard University, Cambridge, Massachusetts 02138, USA}

\author{Abdullah Nasir}
\affiliation{Harvard-MIT Center for Ultracold Atoms, Cambridge, Massachusetts 02138, USA}
\affiliation{Department of Physics, Harvard University, Cambridge, Massachusetts 02138, USA}

\author{Tasuku Ono}
\affiliation{Harvard-MIT Center for Ultracold Atoms, Cambridge, Massachusetts 02138, USA}
\affiliation{Department of Physics, Harvard University, Cambridge, Massachusetts 02138, USA}

\author{Takashi Sakamoto}
\affiliation{Department of Applied Physics, The University of Tokyo, Tokyo 113-8654, Japan}
\affiliation{Harvard-MIT Center for Ultracold Atoms, Cambridge, Massachusetts 02138, USA}
\affiliation{Department of Physics, Harvard University, Cambridge, Massachusetts 02138, USA}

\author{John M. Doyle}
\affiliation{Harvard-MIT Center for Ultracold Atoms, Cambridge, Massachusetts 02138, USA}
\affiliation{Department of Physics, Harvard University, Cambridge, Massachusetts 02138, USA}

\date{\today}

\begin{abstract}

We report a magneto-optical trap of strontium monohydroxide (SrOH) containing \MOTnumber(\MOTnumberUnc) molecules at a temperature of \tempQuarterV(\tempQuarterVUnc)~mK. The lifetime is \tauQuarterV(\tauQuarterVUnc)~ms, which is limited by decay to optically unaddressed vibrational states. This provides the foundation for future sub-Doppler cooling and optical trapping of SrOH, a polyatomic molecule suited for precision searches for physics beyond the Standard Model including new CP violating particles and ultralight dark matter. We also identify important features in this system that guide cooling and trapping of complex and heavy polyatomic molecules into the ultracold regime.

\end{abstract}

\maketitle

\emph{Introduction}---Cold molecules are used for a variety of scientific applications including in quantum information~\cite{demille2002quantum, yelin2006schemes, ni2018dipolar, sawant2020ultracold, wei2011entanglement, yu2019scalable, albert2020robust}, quantum chemistry~\cite{heazlewood2021towards}, and precision searches for physics beyond the Standard Model (BSM)~\cite{kozyryev2017precision, kozyryev2021enhanced, norrgard2019nuclear}. Key to future precision measurements is the effective production and trapping of ultracold molecules to achieve long coherence times. Laser cooling has provided a route to produce ultracold diatomic~\cite{AndereggCaF, WilliamsCaF, BarrySrF, CollopyYO, LimYbF, ZengBaF} and, more recently, polyatomic molecules~\cite{VilasCaOH, KozyryevSrOH, AugenbraunYbOH, MitraCaOCH3, Prehn2016}, which possess features that open a new toolbox for quantum science~\cite{Wall2013, Wall2015, Bohn2019, kozyryev2021enhanced, kozyryev2017precision, Hutzler2020}. For example, a large subset of polyatomic species possess long-lived ``parity doublet'' states arising from rovibrational structures that cannot be found in diatomic molecules. These polyatomic molecule parity doublet states are of particular interest in symmetry-violation searches because they can be mixed in small external electric fields to generate states of molecules that are oriented in opposite directions in the laboratory frame, exhibit closely-matched magnetic moments, and have structural features that are important to the current leading electron electric dipole moment (eEDM) searches~\cite{kozyryev2017precision, AndreevACME, JilaEDM}. Sensitivity to CP-violating new physics through the eEDM scales rapidly with atomic number. As a result, heavy molecules in traps with long coherence times have the potential for improved sensitivity to the eEDM.

Another feature generic to polyatomic molecules (and absent from diatomic molecules) is the existence of multiple vibrational modes (stretching, bending, torsional, etc.). This leads to near-degenerate rovibrational states, which are very rare in diatomic molecules at low energy. These near-degenerate states can be helpful in searches for ultralight dark matter (UDM). For example, as described in Ref.~\onlinecite{kozyryev2021enhanced}, two rovibronic states from different manifolds separated by $\sim$1--10~GHz in SrOH can exhibit different sensitivity to the proton-to-electron mass ratio, $\mu\equiv m_p/m_e$. Precision microwave spectroscopy of the rovibronic transition frequency between these states over time probes $\mu$ variation and, correspondingly, theoretically well-motivated models of dark matter~\cite{arvanitaki2015searching, stadnik2015can, graham2013new, brdar2018fuzzy, banerjee2019coherent, stadnik2016improved, brzeminski2021time, cosme2018scale}.

In this Letter, we report laser slowing and magneto-optical trapping of SrOH molecules in the millikelvin regime. This is the fundamental step toward precision spectroscopy of laser-cooled heavy polyatomic molecules. We observe $\sim$$10^3$ trapped molecules and find MOT characteristics similar to previous molecular MOTs. We also identify important considerations for future work in extending laser cooling methods for polyatomic molecules to even heavier or more complex species. With the results presented here, SrOH is the heaviest molecule with long-lived ($\sim$1~s) parity doublets to be trapped at ultracold temperatures, and represents a favorable platform for future eEDM measurements. In addition, SrOH possesses rovibrational states suitable to probe a broad range of UDM particles via competitive measurements of $\mu$ variation~\cite{kozyryev2021enhanced}.

\emph{Method and apparatus}---Our experimental sequence begins with the production of a cold, slow beam of SrOH molecules. The cryogenic buffer gas beam (CBGB) method is used and has been described previously~\cite{Hutzler2012, Truppe2017}. In short, a metal strontium target inside a \celltemp~K copper cell is ablated by a pulsed Nd:YAG laser. Water vapor is introduced into the cell, reacting with the strontium atoms released in ablation to form SrOH molecules in the gas phase. Helium introduced via a separate fill line thermalizes with the cold walls of the cell and cools the SrOH. Hydrodynamic flow extracts SrOH into a CBGB with typical peak forward velocity of $v_f=$\peakv$\pm$\widthv~m/s. The reaction rate of atoms and water vapor to form SrOH molecules is enhanced by an order of magnitude by directing $\sim$\enhanceP~W of laser light tuned to the $5s^2\,^1S_0$--$5s5p\,^3P_1$ transition of atomic Sr into the cell~\cite{Jadbabaie2020,brazier1985laser}. Approximately \numPulse\ SrOH molecules are produced in the CBGB per ablation pulse.

\begin{figure}[t]
    \centering
    \includegraphics[width = 1\columnwidth]{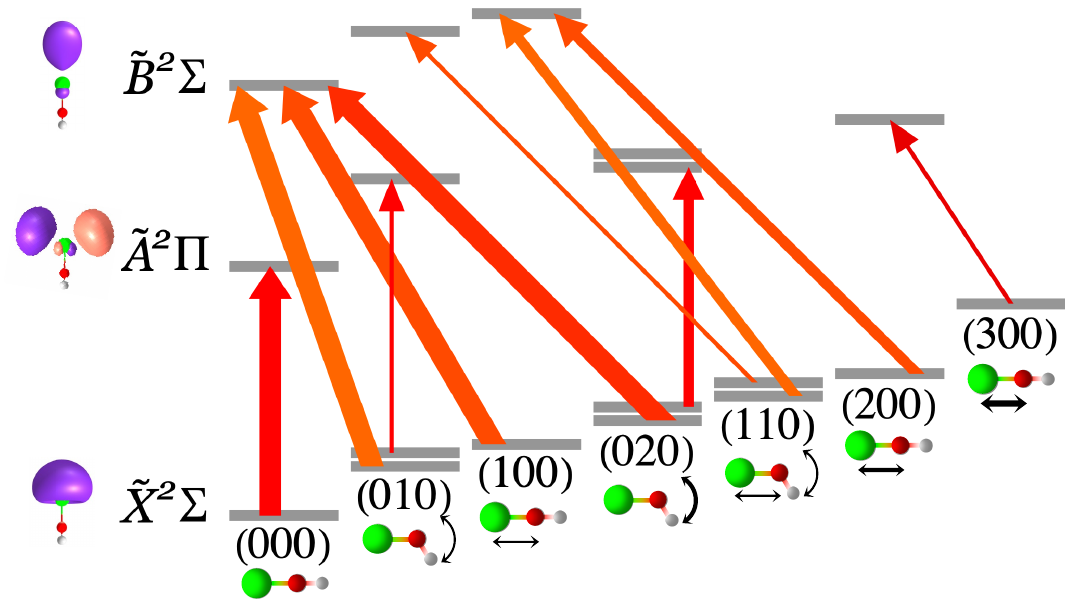}
    \caption{Laser cooling scheme. Vertical position depicts total vibronic energy, up to a fixed overall offset in the $A$ and $B$ energies. Separations between $N=1$ and $N=2$ states within $(010)$ and $(110)$, and between $\ell=0$ and $|\ell|=2$ manifolds of $(020)$, are amplified for clarity. Thin, medium, and thick lines denote slowing lasers with $<$10~mW, $<$100~mW, and $<$1~W, respectively. Illustrations on the left depict the valence orbitals in each electronic state.}
    \label{fig:levels}
\end{figure}

The MOT vacuum chamber is \beamline~cm from the CBGB cell. The molecules, traveling between the cell and MOT, are slowed by lasers counter-propagating to the molecular beam. The slowing lasers are gently focused beyond the cell to mitigate molecular beam divergence. The optical cycle predominantly occurs via the $\X(000;N=1^-)\leftrightarrow\A(000;J=1/2^+)$ transition at $\lambda=688$~nm (the ``main transition''). Here $N$ denotes the rotational quantum number in Hund's case (b), and $J$ is the total angular momentum excluding nuclear spin. Decay to other vibrational states in $\X$ occurs on average after \maincycle\ spontaneous decays, or photon scatters~\cite{LasnerVBRs2022}. By repumping excitations of the Sr--O stretching and Sr--O--H bending vibrational modes (9 repumping lasers total), we increase the average number of photon scatters per molecule, generically denoted the ``photon budget'' $\bar{\gamma}$, to $\bar{\gamma}_{\rm{max}}\approx\cycle$. The rovibronic transitions employed here are shown in Fig.~\ref{fig:levels}. The repumping laser beams are spatially co-aligned with the main transition laser beam, and also pass through the MOT region. In order to maintain a high scattering rate over the entire $\sim$150~MHz Doppler width of the molecular beam, as well as $\sim$110~MHz spin-rotation splittings in $\X$, we employ white-light slowing where each laser is frequency broadened by $\sim$300~MHz via an electro-optic modulator strongly driven with a tank circuit resonant at $\sim$\eomfreq~MHz. The polarization of every laser is switched by a Pockels cell at \slowswitch~MHz to destabilize dark states. The average reduction in velocity per scattered photon is $h/(\lambda m)=5.5$~mm/s, where $m$ is the mass of an SrOH molecule. To stop the beam of SrOH molecules requires scattering $\sim$20000 photons (with about 80\% of molecules lost to unaddressed vibrational dark states given $\bar{\gamma}_{\rm{max}}$). A fraction of the molecules are slowed to below the capture velocity of the MOT, estimated to be $v_f\sim10$~m/s. We have also implemented a chirped slowing scheme by turning off the frequency broadening on the laser addressing the first repumping transition ($\X(100)\leftrightarrow\B(000)$), adding a low-frequency sideband to address the ground state spin-rotation splitting, and ramping the repumping laser frequency over time to maintain resonance with molecules as they are radiatively slowed from their initial velocity to the MOT capture velocity. Chirped slowing increases the number of molecules captured into the MOT by a factor of $\sim$2 compared to white-light slowing.

After the molecules have been decelerated to a sufficiently low velocity, the main transition slowing laser is turned off to avoid pushing molecules out of the trap. The first repumping transition slowing laser is also typically turned off to prevent it from exerting a significant net force on the molecules. When the $\X(100)$--$\B(000)$ slowing laser is turned off, the $\X(100)$ state is repumped in the MOT region via a separate laser beam directed vertically and retroreflected through the MOT center.

The main transition MOT laser, which addresses $\X(000)$--$\A(000)$ but is not frequency broadened, is red-detuned from resonance, contains spin-rotation sidebands (split by 109~MHz), and is spatially split to form the three orthogonal arms of the MOT (the ``MOT beams''). Because the magnetic $g$-factor has opposite sign between the two ground spin-rotation states, the sidebands employ opposite circular polarizations~\cite{VilasCaOH}. The MOT beams have a $1/e^2$ gaussian diameter of 20~mm and powers of up to 10~mW per spin-rotation sideband in each arm. We employ a radio-frequency (RF) MOT with the polarization of the MOT beams switched at \switchfreq~MHz by a Pockels cell to remix magnetic dark states; the electrical current through the MOT coils is driven synchronously. This combination of magnetic fields and laser polarizations maintains a trapping force at all times. The RMS magnetic field gradient along the axial direction is $\sim$\bgrad~G/cm.

\emph{MOT characterization}---We measure the number of trapped molecules ($N_{\rm{mol}}$), their temperature ($T$), the molecular cloud size ($\sigma_0$), and other values of interest under different trapping conditions. The MOT fluorescence is collimated using an in-vacuum lens and imaged on an EMCCD camera to determine $N_{\rm{mol}}$. We detect vibrationally diagonal $\B\leadsto\X$ decays around 611~nm and spectroscopically filter out the light from all laser wavelengths for background-free detection. The number of potentially detectable photons per molecule is approximately the total number of photons scattered from the $\B$ manifold, which is $\sim$650 photons per molecule, as determined by a Markov chain model for the optical cycling process. The photon collection efficiency and camera sensitivity are calibrated separately. Overall, \photonsPerMolecule\ photons are detected per molecule. We determine a typical number of trapped molecules to be $N_{\rm{mol}}=\MOTnumber(\MOTnumberUnc)$, where the uncertainty is dominated by the possible $\sim$30\% error in $\bar{\gamma}_{\rm{max}}$ due to leakage to unidentified vibrational states.

\begin{figure}[t]
    \centering
    \includegraphics[width = \columnwidth]{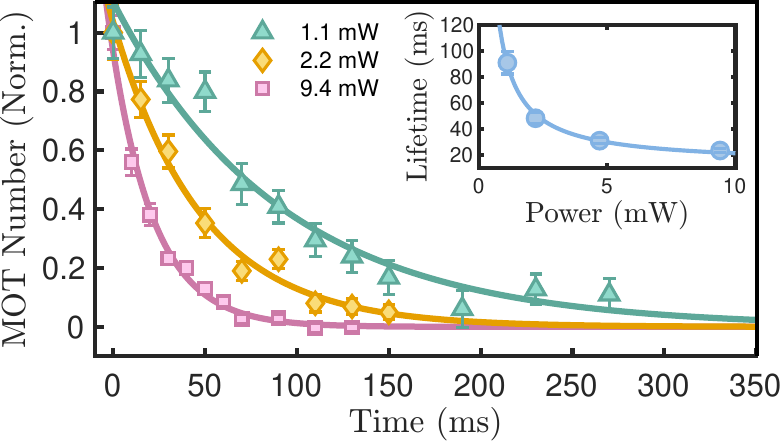}\\
    \includegraphics[width = \columnwidth]{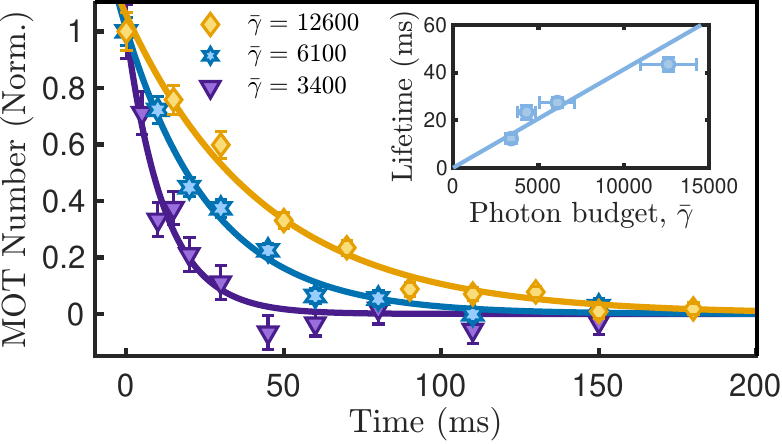}
    \caption{\textbf{Lifetime measurements.} Fluorescence over time and extracted lifetimes. Errorbars denote $1\sigma$ uncertainty. Insets show fitted lifetimes. Top: Dependence on power per MOT beam, with $\tau=\{\tauTwoV(\tauTwoVUnc), \tauOneV(\tauOneVUnc), \tauHalfV(\tauHalfVUnc), \tauQuarterV(\tauQuarterVUnc)\}$ for measurements at \{\PtwoV, \PoneV, \PhalfV, \PquarterV\}~mW. Bottom: Dependence on photon budget $\bar{\gamma}$, with $\tau=\{\tauBOne(\tauBOneUnc), \tauBTwo(\tauBTwoUnc), \tauBThree(\tauBThreeUnc), \tauBFour(\tauBFourUnc)\}$ for $\bar{\gamma}=\{\BOne, \BTwo, \BThree, \BFour\}$, measured with the power of the MOT beams at \PhalfV~mW. Some fluorescence data omitted from the plots for clarity.}
    \label{fig:lifetimes}
\end{figure}

The trapped molecule lifetime $\tau$ is measured by observing molecular fluorescence as a function of time between 35~ms and 300~ms after the end of slowing. We confirm in two ways that $\tau$ is predominantly limited by $\bar{\gamma}$. First, as depicted in Fig.~\ref{fig:lifetimes}, the lifetime ranges from $\tau=\tauQuarterV(\tauQuarterVUnc)$~ms at $\PquarterV$~mW of power per MOT beam to $\tau=\tauTwoV(\tauTwoVUnc)$~ms at $\PtwoV$~mW. Since the scattering rate $R$ is proportional to $P/(P+P_0)$, where $P$ is the power per MOT beam and $P_0$ is the effective saturation power, and $\tau\propto1/R$ is limited by $\bar{\gamma}$, the lifetime follows $\tau(P)=\tau_0(P+P_0)/P$ where $\tau_0$ is the lifetime in the case of a fully saturated scattering rate. At low powers, $P\ll P_0$, we find $\tau\propto1/P$, while at high powers, $\tau \to \tau_0$. We show $\tau(P)$ in Fig.~\ref{fig:lifetimes}, along with a fit to the expected curve resulting in $\tau_0=\tauAsymptote(\tauAsymptoteUnc)$ ms and $P_0=\Psaturate(\PsaturateUnc)$ mW. In our second method of characterizing $\tau$, we are able to show $\tau\propto\bar{\gamma}$ by turning off three different sets of repumping lasers (thus decreasing $\bar{\gamma}$) and measuring $\tau(\bar{\gamma})$; see Fig.~\ref{fig:lifetimes}. For example, with \PhalfV~mW per MOT beam and using all repumpers ($\bar{\gamma}=\bar{\gamma}_{\rm{max}}$), we measure $\tau=\tauRef(\tauRefUnc)$~ms, and by turning off four repumping lasers we set $\bar{\gamma}\approx\bar{\gamma}_{\rm{max}}/4$ to observe $\tau=\tauShortCycle$($\tauShortCycleUnc$)~ms. These measurements show that the lifetime is predominantly determined by the finite photon budget, $\bar{\gamma}$. For each set of repumpers used, $\bar{\gamma}$ and its uncertainty are determined by a Markov chain model with the uncertainty in each vibrational branching fraction propagated.

\begin{figure}[t]
    \centering
    \includegraphics[width = \columnwidth]{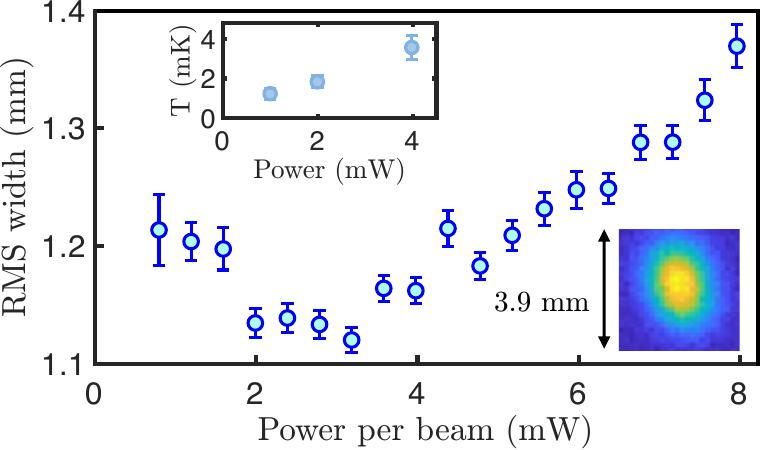}
    \caption{\textbf{Molecule cloud size and temperature.} Geometric mean RMS cloud size $\sigma_0$ and temperature $T$ as a function of power in each MOT beam. The minimum cloud size is \minCloud(\minCloudUnc)~mm and temperatures are $T=\{\tempQuarterV(\tempQuarterVUnc), \tempHalfV(\tempHalfVUnc), \tempOneV(\tempOneVUnc)\}$~mK with \{\PquarterV, \PhalfV, \PoneV\}~mW of power per MOT beam. Error bars show $1\sigma$ uncertainties. A representative image at \PoneV~mW is shown.}
    \label{fig:temperatures}
\end{figure}

The molecule temperature, $T$, is measured using time-of-flight expansion. We image with the MOT beams on resonance and at high power for 5~ms in order to avoid recompressing the MOT during the imaging process. We fit the spatial fluorescence distribution to a 2D gaussian with characteristic sizes $\sigma_{||}$ and $\sigma_\perp$ along the axial and transverse directions, respectively, and extract $T$ along each direction from $\sigma(t)=\sqrt{\sigma_0^2 + (k_B T/m)t^2}$, where $t$ is the free expansion time and $\sigma_0$ is the equilibrium cloud size. The resulting geometric mean temperature is $T=T_{||}^{1/3}T_\perp^{2/3}=\tempQuarterV(\tempQuarterVUnc)$~mK at low power in each MOT beam; $T$ increases with higher MOT beam scattering rates (see Fig.~\ref{fig:temperatures}). We also measure the equilibrium size $\sigma_0$ as a function of MOT beam power and see that it grows as the power increases beyond $\sim$3~mW per beam, as expected.

\begin{figure}[t]
    \centering
    \includegraphics[width = \columnwidth]{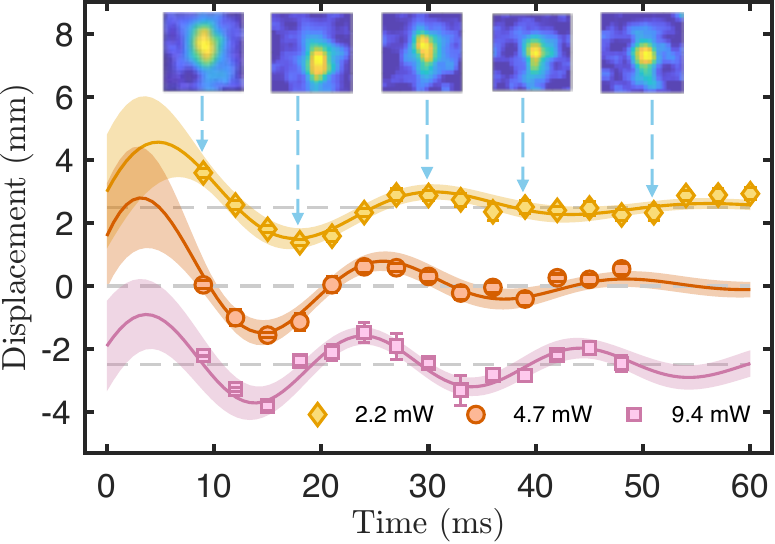}
    \caption{\textbf{Trapped molecule oscillations.} Top: Images of molecule cloud at \PhalfV~mW per beam at several delays after displacement. Bottom: MOT cloud position as a function of time after displacing from the center of the trap by pushing with the $\X(000)$--$\A(000)$ slowing laser, at several values of each MOT beam power (shown offset from each other for clarity). All data are well described by underdamped harmonic oscillators with damping constants $\beta=\{\betaHalfV(\betaHalfVUnc), \betaOneV(\betaOneVUnc), \betaTwoV(\betaTwoVUnc)\}$~s$^{-1}$ and natural oscillation frequencies $\omega=2\pi\times\{\oscHalfV(\oscHalfVUnc), \oscOneV(\oscOneVUnc), \oscTwoV(\oscTwoVUnc)\}$~Hz for \{\PhalfV, \PoneV, \PtwoV\}~mW of power per MOT beam. Errorbars denote 1$\sigma$ uncertainty of data, while fill regions show 95\% confidence interval on the fits.}
    \label{fig:oscillations}
\end{figure}

Finally, we measure the trap oscillation frequency and damping constant by turning on the main transition slowing laser for $\sim$0.5~ms after the MOT has already settled to equilibrium (35~ms after the end of slowing), and observing the resulting damped oscillations for several values of the power per MOT beam. As with measurements of $T$, we image the cloud with the MOT beam on resonance and at high power for 5~ms in order to avoid biasing the cloud position during imaging. The motion near the equilibrium position is described as a damped harmonic oscillator, $a=-\beta v-\omega^2 r$. Under all observed conditions the oscillations are underdamped, and we fit the resulting positions to $r(t)=Ae^{-(\beta/2)t}\cos(\sqrt{\omega^2-(\beta/2)^2}\;t+\phi)$; see Fig.~\ref{fig:oscillations}. Under normal conditions, damping constants are $\beta\sim$100~s$^{-1}$ and the frequency of natural oscillation is $\omega\sim$$2\pi\times45$~Hz.

\emph{Extensions to higher molecular complexity}---The results presented here are comparable to those observed for the first polyatomic molecule MOT of the lighter species, CaOH~\cite{VilasCaOH}, demonstrating the extension of this critical ultracold physics tool to a heavier polyatomic molecule. We observe several features of SrOH in both the optical cycling scheme and the capture of molecules into the MOT that will guide future work with even heavier or more structurally complex species.

Most importantly, we explore the dynamics of loading a MOT of heavy molecules by using MOT beams with $1/e^2$ diameter $d=1$~cm (the same as used to trap CaOH), rather than $d=2$~cm, as used in the results reported in detail above. We find that with $d=1$~cm, although a large fraction of slow molecules experience optical molasses and trapping forces, they do not complete a full oscillation in the trap before exiting the trap region. We refer to molecules with these trajectories as ``nearly trapped.'' The experimental signature of such behavior is a strong fluorescence signal localized at the center of the magnetic field, which, like for fully trapped and damped molecules, depends on using the correct laser polarizations and red detuning but does not persist for more than $\sim$10~ms after the molecular beam is no longer passing through the MOT region. In many respects, this fluorescence signal appears indistinguishable from that arising from ``fully trapped'' molecules (which undergo damped oscillations to the trap center). However, this signal of nearly-trapped molecules can be distinguished from that arising from fully trapped molecules by analyzing the competition between loss rates and loading rates from the tail of the slowed molecular beam pulse. Details of such analysis are described in the Supplemental Material Sec. SM1. By increasing $d$, sufficient length inside the MOT is provided to turn around a large fraction of trappable molecules arriving in the MOT region. Equivalently, increasing $d$ increases the capture velocity. While we do not precisely know the velocity distribution of molecules that reach the MOT region, the velocity distribution of slowed molecules likely scales steeply with velocity such that a modest increase in capture velocity results in a large increase in trapped molecules. Assuming a fixed trapping force, and requiring a fixed capture velocity, the size of the MOT beams are best set to scale linearly with the mass of the species, owing to the lower acceleration of molecules with higher masses. Since SrOH is 1.8$\times$ heavier than CaOH, doubling the beam size can be expected to achieve comparable trapping efficiency.

We also investigate accidental resonances due to the increasing density of states as molecules grow larger. Due to the structural complexity of SrOH, we identify a resonant transition near the main transition, $\X(000;N=1^-)$--$\A(000;J=1/2^+)$ around 435.9682~THz, assigned to the previously unobserved $\X(02^20;N=2^-)\rightarrow\tilde{A}\mu^2\Pi_{1/2}(020;J=3/2^+)$ transition at 435.9690~THz (where $\tilde{A}\mu^2\Pi_{1/2}(020)$ has predominantly $\A(02^00)$ composition). We confirm via scattering rate measurements that this accidental transition plays negligible role in our optical cycle, but stress that for molecules with a larger electronic density of states (e.g., YbOH) or more vibrational modes (e.g., SrNH$_2$ or SrOCH$_3$), accidental excitations of an unwanted transition via some laser in the optical cycle will become more likely. This would result in additional loss channels to rovibronic states. Laser cooling a more complex molecule may require more thorough spectroscopic characterization of rovibronic structure, including of states not directly used for repumping, in order to understand potential off-resonant excitation pathways.

\emph{Conclusion}---We demonstrate and characterize a magneto-optical trap of SrOH molecules, and observe comparable properties to those observed for the lighter isoelectronic molecule CaOH. These results establish the basis for further cooling (e.g., sub-Doppler) and loading into a conservative trap (e.g., an optical diople trap) for precision measurements including the measurement of the electron electric dipole moment and fundamental constant variation. The number of trapped molecules could be further increased in the future by implementing demonstrated and proposed methods such as adding more vibrational repumping lasers to the optical cycle, and transverse laser cooling~\cite{Alauze_2021_YbF_transverse,Langin_2023_improved_loading} of the molecular beam. Combined with recently demonstrated techniques for both diatomic and polyatomic molecules to increase MOT density~\cite{LiBlueMOT, BurauBlueMOT, HallasBlueMOT, JorapurBlueMOT}, substantial numbers of SrOH molecules in an optical dipole trap are feasible. We identify features of trapping heavier or more complex molecules that are likely to be relevant to future work with other species (such as RaOH and SrNH$_2$~\cite{zhang2023relativistic,frenett2024vibrational}), in particular the benefits of large MOT beams to compensate for low accelerations in high-mass molecules.

\begin{acknowledgments}

We are grateful to Yicheng Bao, Christian Hallas, Grace Li, Paige Robichaud and Nathaniel Vilas for valuable technical advice, to Bo Yan for useful discussions regarding the effect of the MOT beam size, and to Andrew Winnicki for assistance with the laser system. We thank Nicholas Hutzler, Tim Steimle, Amar Vutha, and the entire PolyEDM collaboration for helpful input. This work was done at the Center for Ultracold Atoms (an NSF Physics Frontier Center) and supported by Q-SEnSE: Quantum Systems through Entangled Science and Engineering (NSF QLCI Award OMA-2016244), the Heising-Simons Foundation, the Gordon and Betty Moore Foundation, and the Alfred P. Sloan Foundation.

\end{acknowledgments}

\begin{appendix}

\pagebreak
\begin{center}
\textbf{Supplemental Material}
\end{center}

\section*{SM1: Trap loading rate}

We have observed, consistent with observations reported in Ref.~\onlinecite{WilliamsCaF}, that the magneto-optical trap (MOT) of SrOH is loaded from the cryogenic buffer gas beam (CBGB) over an extended time of $\sim$30~ms, when slowing with either chirped-frequency or white-light slowing methods. We characterize the trapped molecule properties, including the number of molecules, by imaging at late times when the loading is nearly complete and the background from the untrappable population of the molecular beam is negligible (thus mitigating potential systematic errors). However, due to the finite lifetime of the trapped molecules, the number observed at these late times is lower than the peak trapped molecule number. Below we describe a method to precisely characterize the trap loading rate over time and infer the peak molecule number from measurements at later times when the trapped number has partially decayed.

\begin{figure}[t]
    \centering
    \includegraphics[width = \columnwidth]{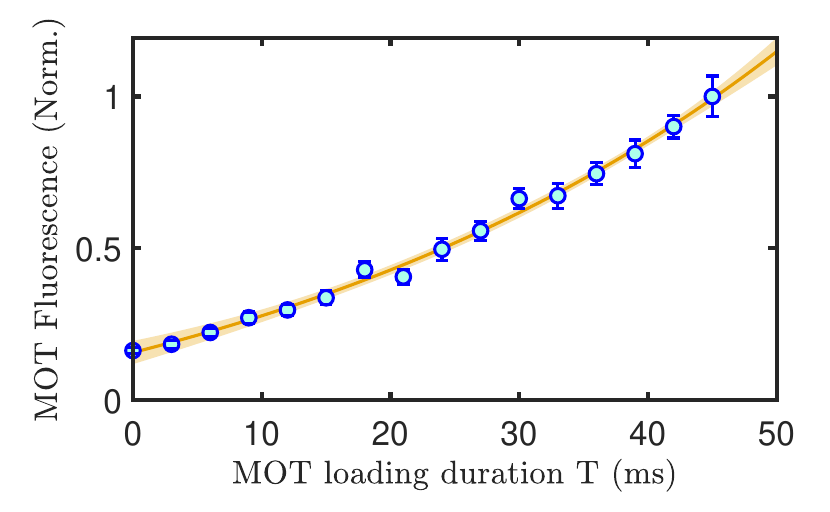}
    \caption{\textbf{MOT fluorescence as a function of loading duration.} Fluorescence measured on a camera exposure beginning at $t = 45$ ms. The fit to Eq. ~\ref{eq:fluorescenceT} is shown in yellow. Errorbars denote 1$\sigma$ uncertainty of data, while fill regions show 95\% confidence interval on the fit. The loading decay time extracted from the fit is $\tau_{\rm{load}}=17.7(1.1)$ ms, while the trapped molecule lifetime was fixed to 29.1(1.7)~ms by independent measurements under the same experimental conditions.}
    \label{fig:Extended_loading}
\end{figure}

We begin by assuming that the MOT loading rate is an exponentially decreasing function of time, $r_{\rm{load}}(t)=r_{\rm{load}}^0 \exp(-t/\tau_{\rm{load}})$. Fits to experimental MOT fluorescence data under different conditions that probe the loading rate, discussed in detail below, validate this assumption as shown in Fig.~\ref{fig:Extended_loading}. We let $t=0$ be the time at which the main transition slowing laser turns off. We confirm that the total number of trapped molecules does not depend on whether the MOT beams turn on at $t<0$ or at $t=0$, and thus infer that $r_{\rm{load}}(t)\approx0$ for $t<0$. This is expected because the slowing lasers provide a large pushing force that overwhelms the trapping force of the MOT and pushes molecules back toward the cryogenic beam source, precluding trap loading~\cite{WilliamsCaF}.

A molecule loaded into the MOT at some time $t_0$ has survival probability at a later time $t_1$ given by $p_{\rm{survive}}(t_1-t_0)=\exp[-(t_1-t_0)/\tau_{\rm{MOT}}]$, where $\tau_{\rm{MOT}}$ is the trapped molecule lifetime. Then the number of molecules in the MOT at time $t$ is given by $N_{\rm{MOT}}(t)=(r_{\rm{load}} * p_{\rm{survive}})(t)$, where $*$ denotes a convolution. This reduces to:

\begin{multline}
N_{\rm{MOT}}(t)=r_{\rm{load}}^0\frac{\tau_{\rm{MOT}} \, \tau_{\rm{load}}}{\tau_{\rm{MOT}} - \tau_{\rm{load}}}\\
\times(\exp[-t/\tau_{\rm{MOT}}]-\exp[-t/\tau_{\rm{load}}]).
\label{eq:MOTnum}
\end{multline}

It follows that the maximum number in the MOT occurs at the time
\begin{equation}
t_{\rm{max}}=\frac{\tau_{\rm{MOT}}\,\tau_{\rm{load}}}{\tau_{\rm{MOT}}-\tau_{\rm{load}}}\ln\left(\frac{\tau_{\rm{MOT}}}{\tau_{\rm{load}}}\right).\label{eq:tmax}
\end{equation}
The trapped molecule number necessarily reaches its peak when the loading rate and decay rate of molecules due to finite trap lifetime balance each other. Therefore, at $t=t_{\rm{max}}$ there must be a non-negligible loading rate of new molecules entering the trap, and interpretation of measurements is not trivial. We instead opt to measure the trapped molecule number at late times, $t\gtrsim2\tau_{\rm{load}}$, when the fraction of trappable molecules that have yet to be loaded is small.

In order to infer the peak number in the MOT from the number observed at a later time $t$, it is necessary to measure $\tau_{\rm{load}}$. We achieve this by observing the number of molecules loaded as a function of the time $t_0$ when the MOT beams turn on (since no molecules can be loaded with the MOT beams off). In particular, let $T\equiv t-t_0$ be the loading duration, where $t$ is the time at which the MOT number is measured. Then the number of molecules in the MOT at time $t$, provided loading begins at time $t_0$, is obtained from Eq.~\ref{eq:MOTnum} with the substitutions $t\rightarrow t-t_0$ and $r_{\rm{load}}^0 \rightarrow r_{\rm{load}}^0 \exp(-t_0/\tau_{\rm{load}})$. The resulting expression simplifies to:
\begin{multline}
N_{\rm{MOT}}(T;t)=r_{\rm{load}}^0\frac{\tau_{\rm{MOT}} \, \tau_{\rm{load}}}{\tau_{\rm{MOT}} - \tau_{\rm{load}}}\exp(-t/\tau_{\rm{load}})\\
\times(\exp[-T/\tau_{\rm{MOT}}]\exp[+T/\tau_{\rm{load}}]-1).
\label{eq:MOTnumT}
\end{multline}

Experimentally, we fix the MOT number measurement time $t$ and vary the loading duration $T$. Then the observed fluorescence signal on a camera exposure, $F$, is proportional to the RHS of Eq.~\ref{eq:MOTnumT}, up to an overall offset due to fluorescence from molecules loaded after time $t$:
\begin{equation}
F(T)=A(\exp[-T/\tau_{\rm{MOT}}]\exp[+T/\tau_{\rm{load}}]-1)+b.
\label{eq:fluorescenceT}
\end{equation}

We measure $\tau_{\rm{MOT}}$ as described in the main text, imaging only after the loading rate is negligible. We then determine $\tau_{\rm{load}}$ by fixing the imaging time of the MOT to $t\approx45$~ms after the end of slowing and varying the duration $T$ over which the MOT beams are on before imaging. Under all MOT conditions, the resulting fluorescence as a function of $T$ fits well to Eq.~\ref{eq:fluorescenceT} and we consistently measure $\tau_{\rm{load}}\approx15$--20~ms from the best-fit curves. Example fluorescence signals taken as a function of the loading time $T$, along with a fit to Eq.~\ref{eq:fluorescenceT}, is shown in Fig.~\ref{fig:Extended_loading}.

From the known values of $\tau_{\rm{MOT}}$ and $\tau_{\rm{load}}$ as determined above, and the number of molecules $N_{\rm{MOT}}(t)$ measured in the MOT at some late time $t\gtrsim2\,\tau_{\rm{load}}$, we can then use Eq.~\ref{eq:MOTnum} to find the peak number of trapped molecules, $N_{\rm{MOT}}(t_{\rm{max}})$. Trapped molecule numbers reported in the main text take into account the scaling $N_{\rm{MOT}}(t_{\rm{max}})\approx 1.4 N_{\rm{MOT}}(t)$.

As discussed above, a detailed understanding and measurement of the time-dependent molecular loading rate into the MOT is necessary to calibrate the peak molecule number. In addition, our model of the extended MOT loading is useful to diagnose MOT performance under conditions where a significant fraction of fluorescence signal arises from ``nearly-trapped'' molecules that never settle into the trap center (e.g., those with initial velocities slightly above the MOT capture velocity). In this regime, the fluorescence signal at late times behaves as $N(t)\propto\exp(-t/\tau)$, with an apparent lifetime $\tau\approx\tau_{\rm{load}}\sim 15$--20~ms. This apparent lifetime $\tau$ will not depend in the expected way on experimental parameters (e.g., laser powers) since it is determined by $\tau_{\rm{load}}$ rather than $\tau_{\rm{MOT}}$, and in particular it may be highly insensitive to changes in experimental conditions. Early observations of the SrOH MOT were well-described by this behavior. Increasing the MOT beam diameters to $d=2$~cm increased the MOT capture velocity sufficiently to allow the trapped molecule fluorescence to dominate over the signal originating from ``nearly-trapped'' molecules. The molecular fluorescence in this regime then reacted to changes in the MOT parameters as expected. We note that similar observations would also occur in situations where $\tau_{\rm{MOT}}\lesssim\tau_{\rm{load}}$, e.g., if the photon budget were too small to support extended photon cycling.

\section*{SM2: Repumper spectroscopy}

Though many rovibrational states in the optical cycle had previously been observed, there were six that had not been identified. In this supplement, we discuss how each of these states was characterized to $\sim$10~MHz resolution using various spectroscopic techniques. Four of the states were identified using the apparatus described in Ref.~\onlinecite{LasnerVBRs2022}, and the remaining two in the apparatus described in the main text used for the slowing and trapping of SrOH. The state energies of the laser-cooling levels can be found in Table~\ref{tab:observed_states}. 

\subsection*{Dispersed Laser-Induced Fluorescence Spectroscopy}

We briefly reiterate the main aspects of the spectroscopy apparatus described in Ref.~\onlinecite{LasnerVBRs2022}. The molecular source is a cryogenic buffer gas beam source (CBGB). The molecules are produced in a copper cell held at $\sim$8~K via laser ablation of a strontium metal target in the presence of $\sim$300~K water vapor and helium buffer gas thermalized with the cell. The SrOH molecules formed from the water vapor and ablated metal are thermalized with the cell by collisions with helium. The cold molecules are then entrained in a beam of helium that exits the cell through a 7~mm aperture. About 2.5~cm downstream, the molecular beam is transversely intersected by a laser beam (where the laser source and addressed transition depend on the specific spectroscopic search). When the light is resonant with a transition, the molecules are excited to an electronic excited state. Fluorescence from the subsequent decay is collected and collimated by a 50.8~mm diameter lens in the cryogenic chamber, subsequently propagating out of the beam source and into a Czerny-Turner style grating spectrometer that disperses the fluoresced light. The wavelength resolution of the dispersion is controlled by an adjustable slit aperture at the input to the spectrometer. This dispersed laser-induced fluorescence (DLIF) is imaged in $40$~nm spans by either an EMCCD camera or gated ICCD camera, for CW and pulsed laser excitation, respectively. The wavelength range is tuneable over $\sim$300~nm by adjusting the angle of the diffraction grating. The wavelength axis of the collected fluorescence is calibrated using known transitions.

Unless indicated otherwise, all transitions address $J^P=1/2^+$ in the excited vibronic state, where $J$ is the angular momentum excluding nuclear spin and $P$ is parity.

\begin{table}[t]
    \centering
    \begin{tabular}{c|c}
    State & Energy ($\wn$) \\
    \hline \hline
    $\B(010;J=1/2^+)$  & 16778.74\\
    $\A(200;J=1/2^+)$ & 15621.93\\
    $\tilde{A}\mu^2\Pi_{1/2}(020;J=1/2^+)$ & 15275.84\\
    \At$\kappa^2\Sigma(010;J = 1/2^+)$ & 15191.15\\
    $\X(300;N=1)$ & 1566.86\\
    $\X(110;N=2)$ & 884.74\\
    $\X(110;N=1)$ & 883.75\\
    \end{tabular}
    \caption{Energies of the previously unobserved rovibronic states used for optical cycling of SrOH, identified with a combination of DLIF and depletion-revival spectroscopy. The absolute uncertainty is conservatively estimated to be $\lesssim$1~GHz by comparison of wavemeter measurements of resonance frequencies for transitions previously observed in our apparatus to absolute frequencies for the same transitions reported in the spectroscopic literature.}
    \label{tab:observed_states}
\end{table}

\subsubsection*{High-resolution $\B(010)$ spectroscopy}

The $\B(010)$ state is used to repump the $\X(110;N=2)$ state. The excited vibronic manifold had previously been observed~\cite{presunka1993high}, but only for higher rotational states. It was thus necessary to identify the energy of the $\B(010)$ state at high resolution. 

To find this state, a CW dye laser was used to drive rotational lines on the $\X(000)$--$\B(010)$ vibronic transition. The $\B(010)$ state remains vibrationally diagonal like other low-lying vibrational states in the $\A$ and $\B$ manifolds, and so upon excitation predominantly decays to $\X(010)$. The frequencies of the excitation scatter and decay light differ by $\sim$360 $\wn$, a splitting easily resolvable on the spectrometer, even with a low-resolution widely-opened slit.

Since the $\X(000;N=1)$--$\B(010)$ transition had not been observed, we first identified the $\X(000;N=3)$--$\B(010;N=3)$ line in our apparatus, which is listed in Ref.~\onlinecite{presunka1993high}. We found a $\sim$300~MHz offset between our frequency reference and the published value, well within the uncertainty of the absolute accuracy of the wavemeter used. 

To identify the $\X(000;N=1)$--$\B(010)$ transition, we calculated the energy using Eqs. 3 and 4 in Ref.~\onlinecite{presunka1993high} and the constants in the previous work, accounting for the observed frequency offset. The line was identified almost exactly where this calculation predicted, with strong fluorescence observed over a few hundred MHz. The width is attributed to both Doppler broadening and the partially resolved ground state spin-rotation (SR) splitting of $\sim$110~MHz.

\subsubsection*{Low- and high-resolution $\A(200)$ spectroscopy}

The $\A(200)$ state is used to repump population out of $\X(300)$. This vibrational manifold had previously never been observed. We initially performed pulsed-dye laser DLIF spectroscopy, with a $\sim$0.1$\wn$ laser linewidth and high intensity capable of driving comparatively weak transitions. The dye laser is a HyperDye-300 pumped by a 10 ns, $\sim$100~mJ YAG, producing more than 2~mJ/pulse of output energy. The dispersed fluorescence was collected by an ICCD, with the intensifier on the camera gated a few ns after the excitation pulse was fired to avoid saturating the sensor.

We estimated the position of the $\A(200)$ manifold from the $\A(100)$ position and the ground state $x_{11}$ anharmonic constant. We then centered the dye laser in the vicinity of this prediction, and scanned the frequency in $\sim$0.1$\wn$ steps. By recording the fluorescence spectrum as a function of excitation frequency, we deduced the nature of the transitions we were driving by the separation between the excitation and decay light frequencies, and additionally by the pattern of decay peaks. In particular, the $\A(200)$ manifold was identifiable by a strong decay $\sim$1000$\wn$ higher energy that the excitation light, and weaker decays to both red- and blue-degraded bands $\sim$500$\wn$ from this dominant decay, indicating stretching character with a few percent of decays to $\Delta v_1 = \pm1$ vibrational levels. The state was quickly discovered.

Low resolution ($\sim$1\%) vibrational fractions (VBFs) were measured with the pulsed dye laser frequency held at the low-frequency edge of the feature. These measurements suggest an $80\%$ diagonal decay to $\X(200)$, an $11\%$ fraction to $\X(300)$, and an $8\%$ fraction to $\X(100)$. Less than $2\%$ of the population decays to $\X(400)$, with lower probability decays below the measurement resolution. These VBF measurements were important for our later utilization of the state, since it confirmed new vibrational decay channels would not arise as long as the $\A(200)$ was populated infrequently ($\lesssim$ 50 times). 

To find the rotational photon cycling line, we also conducted high-resolution narrow-band CW DLIF spectroscopy on the $\X(000)$--$\A(200)$ transition, similar to the previously discussed work identifying the $\B(010)$ repumping state. After locating a few rotational features, the spectrum was fit using the PGopher program. The resulting fit predicted the position of the optical cycling transition, which we confirmed spectroscopically by tracing rotational and spin-rotation spacings across several rotational features. The energy of the $\A(200)$ excited state can be found in Table~\ref{tab:observed_states}.

\subsubsection*{Low- and high-resolution $\X(110)$ spectroscopy}

The decay from the $\A$ and $\B$ manifolds to $\X(110)$ had been observed at low resolution in Ref.~\onlinecite{LasnerVBRs2022}. The observation made clear that it was necessary to repump from the $\X(110)$ manifold to sufficiently photon cycle for a MOT, but the data was not of high enough resolution to locate the repumping transition. Additionally, given that it is the lowest vibrational combination mode, the size of possible anharmonic contributions to the energy were unknown. 

To locate the position of the state more accurately, we started by driving the $\X(000)$--$\B(010)$ transition. Averaging of the dispersed fluorescence revealed a small decay from the excited state, consistent with $\Delta v_1 = 1$ (to $\X(110)$). The axis of this spectral range was calibrated to $\sim$1$\wn$ using several known laser frequencies. This calibration allowed extraction of the center of the presumed decay to $\X(110)$ to similar uncertainty. 

Similar to the $\A(200)$ state, no rotational structure of the level had been previously identified. Scanning a CW laser over $\sim$1$\wn$ around the $\X(110)$--$\B(010)$ best-guess origin revealed several rotational features. Fitting of this decay spectrum was again assisted by PGopher to assign the lines. The resulting fit successfully predicted the energy of the $\X(110;N=1)$--$\B(010)$ repumping transition. The $\X(110;N = 2)$ state was then identified by adjusting the laser frequency by the expected rotational energy level difference, confirming the assignment. The energy of the $\X(110;N=1-2)$ states can be found in Table~\ref{tab:observed_states}.

We note that the spin-rotation splitting observed in the $\X(110;N=1)$ manifold, as well as in the $\X(010;N=1)$ manifold, is around 110~MHz, comparable to that observed for non-bending vibrational states. However, the spin-rotation splitting in the $N=1$ level of an $\ell=1$ vibrational manifold is expected to be half as large as in $\ell=0$ manifolds (see Ref.~\onlinecite{merer1971rotational} for relevant matrix elements). We have also seen the spin-rotation splittings for low-$J$ states in $\B(010)$ to be well-described by the formulas applicable to $\ell=0$ manifolds. Measurements in our group of the CaOH spin-rotation splittings \emph{do} show the expected distinction between $\ell=0$ and $\ell=1$ manifolds. We have not identified the underlying reason for the comparable size of the spin-rotation splittings in $\ell=0$ and $\ell=1$ manifolds in SrOH.

\subsection*{Depletion-Revival Spectroscopy}

The DLIF spectroscopy above relied on the ability to drive a relatively strong transition out of a reasonably populated ground state (roughly speaking, $\X(000)$, $\X(010)$, or $\X(100)$, with $X(110)$ and $\X(020)$ having small but detectable populations). Combined with driving vibrationally off-diagonal transitions, this combination ensured a strong fluorescence signal whenever the excitation laser was resonant with a transition. Spectroscopy of the $\X(300)$ ground states and \At($02^00)$ excited states using the same approach is not ideal: $\X(300)$ is expected to barely be naturally populated after entrainment in the buffer gas beam, and $\A(02^00)$ is not well-coupled to any highly-populated ground states.

The apparatus used in the following work was the MOT apparatus described in the main text. Reiterating briefly, a CBGB beam of SrOH molecules traverses a \beamline~cm long beam line. Any or all of the optical cycling lasers can counter-propagate against the molecular beam (with frequency broadening). Molecules are detected in the MOT chamber at the end of the beam line with different detection schemes in each case, with fluorescence collimated by an in-vacuum lens and collected by a PMT outside of the vacuum chamber.

The spectroscopy of states in this apparatus was done via depletion-revival spectroscopy. In this class of schemes, a laser first pumps population through some transition that empties a specific ground state or set of ground states. The newly populated states can then be repumped through an excited state, which can ``revive'' the population in a detection scheme. This method is especially effective when optical cycling, where including specific repumpers can be used to cycle effectively into a higher vibrational ground state, and/or make evident the return or disappearance of the population from the optical cycle. The experimental arrangements vary among the several spectroscopic searches and are each discussed in the corresponding subsections below.

The $\A\kappa(010)$ repumping state was also identified via this method, despite nominally being compatible with DLIF searches as well.

\subsubsection*{High-resolution $\X(300)$ spectroscopy}

The $\X(300;N=1)$ level, one of the least populated states in the optical cycle, had not been previously observed. However, unlike the $\X(110)$ manifold, the dominant anharmonic contributions to the stretching vibrational mode energies were measured in previous work on the $\X(100)$ and $\X(200)$ manifolds. This allowed prediction of the $\X(300;N=1)$ energy to within a few GHz using standard rovibrational energy formulas and previously measured constants. 

To populate the state, the $\X(000)$--$\A(200)$ line was driven. From this transition, about 10$\%$ of the molecules decay into $\X(300)$ in a single excitation. To increase the population further, the $\X(000)$--$\A(000)$, $\X(100)$--$\B(000)$, and $\X(200)$--$\B(100)$ transitions in the optical cycle were also driven by white-light broadened slowing light, which depleted the lower-lying stretching modes and increased the $\X(300)$ population. 

A Ti:Sapph probe laser transversely intersected the molecular beam in a detection chamber downstream, and was scanned near the estimated $\X(300)$--$\A(100)$ transition. The depletion lasers were kept on continuously. When the scanning laser was resonant with the photon cycling line, the population from the $\X(300)$ state was returned predominantly to the $\X(100)$ state, which had been depleted of natural population. The $\X(000)$--$\A(000)$ and $\X(100)$--$\B(000)$ lasers could then cycle the revived population $\sim$400 times before loss. A PMT with spectroscopic filters collected 611~nm fluorescence from $\B(000)\leadsto \X(000)$ decays, allowing background-free detection. In this way, transitions in the $\X(300)$--$\A(000)$ band were identified to high resolution. The energy of the $\X(300;N=1)$ level is recorded in Table~\ref{tab:observed_states}, which enabled driving of the actual repumping transition through $\A(200)$.

\subsubsection*{Low- and high-resolution $\tilde{A}\mu^2\Pi_{1/2}(020)$ spectroscopy}

The $\A(02^00)$ and $\tilde{A}^2\Pi_{3/2}(02^20)$ states couple to each other via second-order Renner-Teller interactions to produce the $\tilde{A}\mu^2\Pi_{1/2}(020)$ state (with predominantly $\A(02^00)$ composition) and $\tilde{A}\kappa^2\Pi_{1/2}(020)$ state (with predominantly $\tilde{A}^2\Pi_{3/2}(02^20)$ composition). This coupling allows us to repump $\X(02^20)$ with non-negligible strength through $\tilde{A}\mu^2\Pi_{1/2}(020)$, which in turn predominantly decays to the $\X(02^00)$ state.

The $\tilde{A}\mu^2\Pi_{1/2}(020)$ manifold had not been previously observed, and the position was estimated from low-resolution DLIF taken with a pulsed dye laser. The high-resolution spectroscopy was done using a depletion-revival method by populating the $\X(02^00)$ state and subsequently driving the strong $\X(02^00)$--$\tilde{A}\mu^2\Pi_{1/2}(020)$ transition to transfer some fraction of population to unobserved vibrational states (predominantly $\X(12^00)$).

To populate $\X(02^00)$, the earlier lasers in the optical cycle were pulsed for $\gtrsim$10~ms counter-propagating to the molecular beam. A Ti:Sapph probe laser transversely intersected the molecular beam downstream, scanning over the best estimate for the $\X(02^00)$--$\tilde{A}\mu^2\Pi_{1/2}(020)$ transition frequency. When the probe laser was resonant with the transition, the population that had accumulated in $\X(02^00)$ was emptied out. The laser ordinarily used to repump $\X(02^00)$ in our optical cycle (via $\B(000)$) was then used as a detection laser downstream, with fluorescence from the molecules collected by a PMT. When the spectroscopy laser was off-resonant, the detection laser induced fluorescence from the population in the $\X(02^00)$ state. When the resonance condition was met, the induced fluorescence was much smaller since the spectroscopy laser depleted the detected state. Using this method, several rotational features, including the photon-cycling line, were identified successfully through the pattern of rotational and spin-rotation splittings. The energy of the $\tilde{A}\mu^2\Pi_{1/2}(020)$ repumping state is noted in Table~\ref{tab:observed_states}.

\subsubsection*{High-resolution \At$\kappa^2\Sigma$$(010)$ spectroscopy}

The \At$\kappa^2\Sigma(010)$ state is used to repump $\X(010;N=2)$. Originally, the $\X(010;N=2)$ ground state was repumped through $\B(000)$, but this nominally forbidden transition was not strong enough to drive with the desired scattering rate given our laser system. Driving through the \At$\kappa^2\Sigma(010)$ state is necessarily much stronger (since it is a vibrationally diagonal line) and also accessible via diode laser at 674~nm. The vibronic \At(010) levels had been observed previously in Ref.~\onlinecite{Presunka1994}, but the low $J$ states had not been recorded in the $\mu^2\Sigma$ manifold. Nevertheless, the spectroscopic constants and the matrix elements therein allowed calculation of the repumping transition $\X(010;N=2)$--\At$\kappa^2\Sigma(010)$. The $\X(010;N=2)$--\At$\mu^2\Sigma(010)$ transition was also considered for repumping, but ultimately not chosen due to its less convenient wavelength to integrate into the rest of our optical system.

To locate the \At$\kappa^2\Sigma(010;J = 1/2^+)$ and \At$\mu^2\Sigma(010;J = 1/2^+)$ states, a similar method to the $\X(300)$ search was used. All lasers in the full optical cycling scheme were sent counterpropagating to the molecular beam to deplete as much of the population into $\X(010;N=2)$ as possible. Although this state is not strongly populated in the optical cycle, it is also low-lying enough to have non-negligible natural population originating from the CBGB. The cycling thus served both to increase $\X(010;N=2)$ population and to empty the rest of the ground states in the optical cycle for higher measurement contrast. A probe laser then intersected the molecular beam as it scanned near the calculated resonances. First, a Ti:Sapph laser was used to probe the $\X(010)$--\At$\mu\Sigma(010)$ transition. While the laser frequency was scanned, the fluorescence from the excited population was collected by a PMT. When the transition was reached, the fluorescence increase was mapped over the peak.

After this state was found, the process was repeated searching for the \At$\kappa^2\Sigma(010)$ vibronic manifold. Given the confirmation of the low-N transition to the other excited $\Sigma$ vibronic manifold, this transition was predicted to within a few hundred MHz, and the transition was quickly observed. The energy of the \At$\kappa^2\Sigma(010;J=1/2^+)$ state can be found in Table~\ref{tab:observed_states}.

\section*{SM3: Design and characterization of the repumping scheme}

An optimal repumping scheme returns molecules quickly to the vibronic ground state to maintain a high photon scattering rate and uses technologically accessible laser wavelengths and powers. In SrOH, a convenient way to generally achieve these goals is by repumping through the strongest available transition that reduces the total number of vibrational excitation quanta (e.g., $\X(300)$--$\A(200)$), even if a weaker transition is available that reduces the number of vibrational excitation quanta by a larger amount (e.g., the relatively weak $\X(300)$--$\A(100)$ transition). The repumping transitions that reduce one stretching excitation quantum through the $B$ state (e.g., $\X(100)$--$\B(000)$) are at an exceptionally convenient wavelength around 630~nm, where high-power, stable sum-frequency generation (SFG) lasers based on a combination of Yb- and Er-doped fiber amplifiers are available. The equivalent transitions through the $A$ manifold occur around 710~nm, a more difficult wavelength to produce in SFGs at high power.

In this supplement, we provide technical details for the construction of our repumping laser systems and an overview of how each required laser power was determined. Additionally, we discuss how molecular perturbations affect the strength of late repumping transitions (i.e., those that become relevant in the optical cycle after thousands of photon scatters).

\subsection*{Laser systems} \label{sec:techDetails}

\begin{table*}[t]
%\resizebox{\textwidth}{!}{%
\begin{tabular}{|c|c|c|c|c|c|}
\hline
\textbf{Ground State} & \textbf{Excited State} & \textbf{Wavelength} & \textbf{Source} & \textbf{$f_i$} & \textbf{Power} \\ \hline
$X(000)$ & $A(000)$ & 687.6 nm & Tm+Yb & $9.55 \times 10^{-1}$ & 800 mW \\ \hline
$X(100)$ & $B(000)$ & 630.9 nm & Er+Yb & $4.21 \times 10^{-2}$ & 800 mW \\ \hline
$X(200)$ & $B(100)$ & 630.5 nm & Er+Yb & $1.58 \times 10^{-3}$ & 60 mW \\ \hline
$X(010;N=1)$ & $B(000)$ & 624.5 nm & Er+Yb & $4.52 \times 10^{-4}$ & 250 mW \\ \hline
$X(02^00)$ & $B(000)$ & 638.0 nm & Er+Yb & $3.72 \times 10^{-4}$ & 200 mW \\ \hline
$X(02^20)$ & $A\mu^2\Pi_{1/2}(020)$ & 687.7 nm & Ti:Sapph & $2.53 \times 10^{-4}$ & 50 mW \\ \hline
$X(010;N=2)$ & $A\kappa^2\Sigma(010)$ & 674.5 nm & ECDL & $8.95 \times 10^{-5}$ & 5 mW \\ \hline
$X(300)$ & $A(200)$ & 711.5 nm & ECDL & $7.48 \times 10^{-5}$ & 6 mW \\ \hline
$X(110;N=1)$ & $B(100)$ & 623.9 nm & Er+Yb & $6.82 \times 10^{-5} $ & 25 mW \\ \hline
$X(110;N=2)$ & $B(010)$ & 629.2 nm & Er+Yb & $1.70 \times 10^{-5}$ & 10 mW \\ \hline
\end{tabular}%
%}
\caption{Summary of laser systems for optical cycling in SrOH. The fraction of decays within the optical cycle that populate state $i$ is denoted $f_i$, as discussed in the text. Powers shown are typical powers in the slowing beams. For lasers generated by SFGs, the doping of the fiber amplifiers which produce the IR light is noted.}
\label{tab:lasers-table}
\end{table*}

The majority of our lasers, as listed in Table~\ref{tab:lasers-table}, are sum-frequency generation (SFG) systems, which produce visible light by combining infrared light at two distinct wavelengths in a nonlinear crystal. These systems tend to be robust, high-power sources of laser light requiring very little regular maintenance. 

The main transition laser, $\X(000)$--$\A(000)$ at 687.6~nm, is generated by a turnkey commercial SFG system (Precilasers) that sums infrared light produced by Tm- and Yb-doped fiber amplifiers.

Our other SFG systems are produced by summing infrared light from Er- and Yb-doped fiber amplifiers. The sum frequency generation was implemented in home-built optical systems in order to conveniently accommodate possible adjustments to the required laser system during early development of the experiment. All of these amplifiers are seeded by DFB lasers (Precilasers EFL-1550 for Er-doped amplifiers, and Innolume DFB-XXXX-PM-30-OI-LT for Yb-doped amplifiers, where XXXX denotes the seed wavelength). Each fiber amplifier can output up to 10~W of infrared light.

\begin{figure}
    \centering
    \includegraphics[width=\columnwidth]{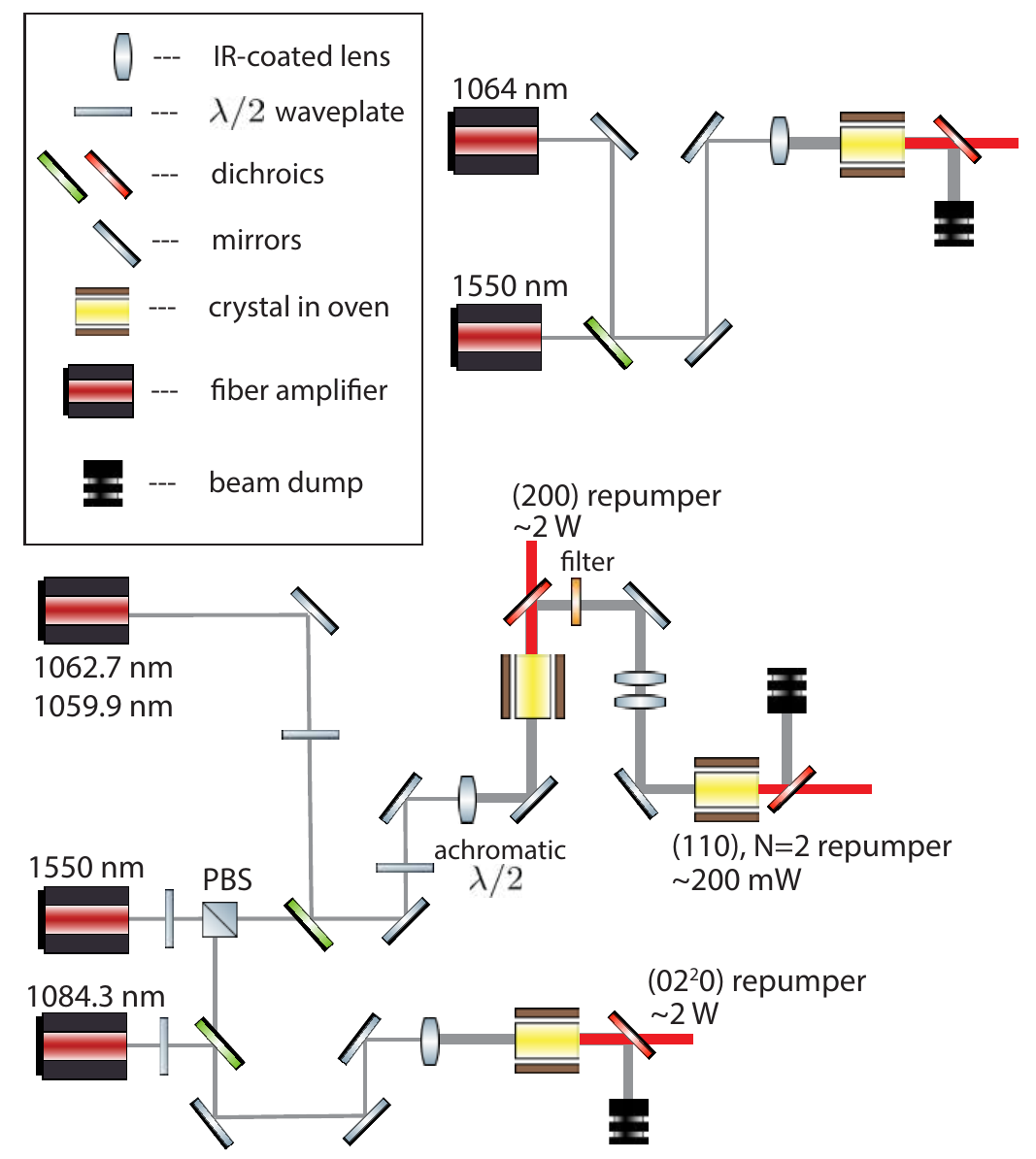}\\
    \caption{Schematic for a sum-frequency generation laser system (top-right) and an example optical path for producing 3 of our repumping laser wavelengths via daisy-chained crystals (bottom).}
    \label{fig:SFGs}
\end{figure}

In each SFG system, the two high-power IR beams are overlapped on a dichroic mirror, rotated to S-polarization, and focused into an MgO-doped periodically poled lithium niobate (MgO:PPLN) crystal mounted in a temperature-stabilized oven; see Fig.~\ref{fig:SFGs}. By adjusting the crystal temperature to achieve a quasi-phase matching condition, visible light can be produced with up to around 7~W of power. The visible wavelengths for these specific systems range from around 624~nm to 638~nm. A more thorough guide to aligning such systems can be found in Appendix~A of Ref.~\onlinecite{Baumthesis}. 

The straightforward SFG system described above is useful when powers $>$~1~W are needed. However, as seen in Table~\ref{tab:lasers-table}, most repumpers do not employ such large optical powers. To use resources more efficiently, the IR power from the one amplifier may be used to generate several visible wavelengths~\cite{Baothesis}. We employ two approaches to re-use the output of an amplifier. First, we can split the output from an amplifier along multiple spatial paths. Second, we can seed an amplifier with multiple low-power fiber lasers so that the high-power output has contributions at all seed wavelengths. An illustration of both approaches, in the same optical system, is shown in Fig.~\ref{fig:SFGs}.

To multi-seed a Yb-doped fiber amplifier, the different 1064~nm seeds are typically combined on 50:50 PM fiber couplers. Provided both wavelengths are within the gain curve of the fiber amplifier, they can be amplified to a combined power of around 10~W. The remainder of the alignment procedure is then similar to the single SFG example and is represented in Fig.~\ref{fig:SFGs}. The combined light enters the first PPLN crystal, where temperature tuning is used to selectively phase-match the 1550~nm with one of the $\sim$1064~nm seeded wavelengths. Then, the produced visible light is separated from the IR and sent to the experiment. The remaining IR light is recollimated and refocused into a second crystal (``daisy-chaining''), whose temperature is tuned to phase-match with a different $\sim$1064~nm wavelength than the first crystal. Due to the decreased 1550~nm power and a worse beam shape, the power output from the second crystal is always lower than the first in our SFG systems. The order in which the light is generated is determined by the power required for each repumping transition.

The temperature tuning allows for good selectivity in generating approximately one wavelength per crystal. However, repumpers addressing two adjacent rotational states ($\sim$30~GHz apart), produced by daisy-chaining the crystals, exhibited interference when combined into a single beam for the slowing experimental work. This interference occured because the rotational spacing is on the same order as the quasi-phase matching bandwidth of the crystal. In this case, both crystals necessarily produced some amounts of each wavelength even when the crystals were kept at different temperatures. The resulting visible beams will then contain identical frequency components, which interfere with each other after being recombined along the same beam path. The relative phase between the components slowly drifted due to small fluctuations in the optical path length difference, resulting in significant power fluctuations along the combined path (e.g., for slowing the molecular beam). For our rotational spacing, we observed interference fluctuations on the order of $\sim$10\%, which decreased our average power to the experiment but did not provide any further experimental challenges or generally preclude us from daisy-chaining crystals spaced by $\sim$30~GHz. It is helpful to optimize the temperatures for the crystals by tuning them with only one seed on at a time; otherwise the temperature that maximizes the total visible output power tends to produce relatively large power at both closely-spaced wavelengths in both crystals (thus resulting in significant interference between the optical paths). Nevertheless, some amount of interference cannot be avoided, without sacrificing power output from the crystals (e.g., by optimizing each crystal temperature to minimize the output at the unwanted wavelength). To force the interference fluctuations to occur significantly faster than the experimental timescale of $\sim$1~s, we drove AOMs, used in each optical path as switches before the repumpers are combined, with RF signals differing in frequency by $>\sim$2~MHz. As a result, the interference between the two paths beats at the AOM drive frequency difference after being combined.

We currently produce the repumpers for $\X(02^00)$, $(200)$, and $(110; N=2)$ by daisy-chaining the crystals, as in Fig.~\ref{fig:SFGs}, and splitting 1550~nm amplifier power. The repumpers for $\X(010; N=1)$ and $X(110; N=1)$ are generated similarly by daisy-chaining crystals.

We lock all of the SFG repumpers to within $\pm$10~MHz (and the main transition laser to within $\pm$1~MHz), by feeding back on the 1064~nm seeds. Thus, these SFG systems are easy to lock and are stable in power over time. When considering experimental power requirements, a fairly large power overhead out of the lasers is necessary to account for inefficiencies due to combining the closely spaced wavelengths as well as decreased fiber coupling efficiencies for beams passing through strongly driven EOMs. However, the SFG systems produce high enough powers stably that the powers listed in Table~\ref{tab:lasers-table} are easily achievable.

\subsection*{Determining Repumping Power Requirements}\label{sec:scatteringRates}

In this section, we describe how to measure scattering rates for each transition in the optical cycle, which is useful for diagnosing issues with repumping lasers before slowing or trapping is achieved. The scattering rate of the main transition is measured to be $\Gamma_0\approx$ 2~MHz. As repumping lasers are added, ideally the scattering rate over the entire optical cycle would remain as close as possible to $\Gamma_0$ to maximize optical forces. One key consideration to maintain a high scattering rate is to avoid repumping vibrationally excited states through the $\A(000)$ state addressed by the main transition laser, since such a repumping transition would decrease the average time spent in $\A(000)$ and thus decrease the overall rate of spontaneous decays. Even so, the overall scattering rate of the scheme is partially limited by how long the molecules spend in excited vibrational states in the ground electronic manifold.

During early steps toward laser slowing, it was important to confirm that each repumper was both at the correct detuning and had sufficient power. Before having any slowing or MOT signal, measuring the scattering rate for each laser was the clearest way to characterize the efficacy of each repumper.

\begin{figure*}
    \centering
    \includegraphics[width=\textwidth]{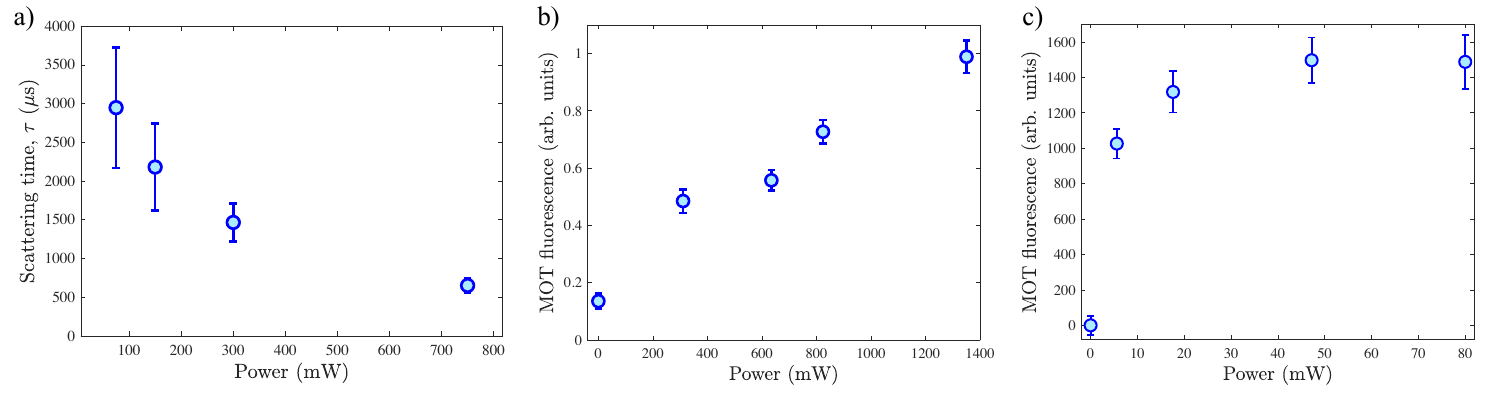}\\
    \caption{Single-photon scattering time and MOT signal as a function $\X(02^20)$ repumping laser power. Subfigures (a) and (b) show power scans for the $\X(02^20) - \A(100)$ transition, and subfigure (c) was taken for $\X(02^20) - A\mu^2\Pi_{1/2}(020)$. The single photon scattering time for the repumper through $\A(100)$ in subfigure a) only achieves a repumping time of 500~$\mu$s at 750~mW. Although this corresponds to a reasonably large value of $\Gamma_{\rm{out}}/\Gamma_{\rm{in}}\sim5$, we note that it is on the order of the $\sim$1~ms repumping timescale where slowing and trapping dynamics occur, which might further limit repumping efficacy. Indications of this limitation can be seen in (b) where even 1.35~W (higher than any other repumping laser power) did not saturate the MOT fluorescence. After switching to repumping $\X(02^20)$ through $A\mu^2\Pi_{1/2}(020)$, the MOT fluorescence was saturated with 40~mW.}
    \label{fig:combined-power-scans}
\end{figure*}

We measured the scattering rate of the main transition laser by turning it on in increasing time increments, pumping molecules out of $\X(000)$. Measurement of the remaining population as a function of laser pumping time, combined with our measured VBFs from $\A(000)$, allowed the inference of the effective scattering rate. We measured the combined scattering rate of the main transition and first repumper ($\X(100)-\B(000)$) with a similar method, again calculating a photon budget for these two ground states from our VBFs to back out the final rate. We measured a combined scattering rate of 2~MHz.

For all later repumpers, the single photon scattering rate (i.e., the inverse of the mean time to scatter a single photon), was measured instead in order to isolate the performance of each individual laser. To measure the single photon scattering rate, we first populated the state of interest, $i$, by turning its repumping laser off in the slowing laser beam. Then, we repumped $i$ for a variable amount of time and finally detected the resulting population returned to the optical cycle. As the state was addressed for longer times, the number of molecules recovered saturated, and we fit the single photon scattering rate to the curve of repumped population vs. repumping time. 

The single photon scattering rate out of an excited vibrational state, $\Gamma_{\rm{out}}^{(i)}$, was compared to the rate we would expect molecules to fall into the state, $\Gamma_{\text{in}}^{(i)} \sim \Gamma_0  f_i,$ where $f_i$ is the branching fraction into state $i$. It is not feasible, nor necessary, for all repumpers to have scattering rates comparable to $\Gamma_0$: naively, one might expect that a repumper's power is sufficient provided $\Gamma_{\rm{in}}\gtrsim\Gamma_{\rm{out}}$. However, we have empirically seen otherwise, as shown in Figures~\ref{fig:combined-power-scans}a and \ref{fig:combined-power-scans}b. In particular, as more ground states are added to an optical cycle, it becomes necessary to pump molecules out of the various excited vibrational ground states increasingly quickly, even if the ground states are not coupled to the same excited states.

To understand the requirements for repumping rates as a function of the number of lasers in an optical cycle, we consider a simplified set of rate equations describing the evolution of the population among ground states,
\begin{equation}\label{eq:rate-eq}
    \frac{\mathrm{d} p_i}{\mathrm{d} t} = -p_i (1 - f_i) \Gamma_i + \sum_{i \neq j} p_j  f_i \Gamma_j,
\end{equation}
where $p_i$ is the fraction of molecular population in state $i$, $f_i$ is the branching fraction into that state (assumed to be independent of the previously occupied state, and normalized so that $\sum_i f_i=1$ for states in the optical cycle), and $\Gamma_i$ is the scattering rate for the laser that addresses $i$. Representative values of $f_i$ for the SrOH optical cycle, determined by a Markov chain that uses known or estimated vibrational branching fractions from each employed excited state, are listed in Table~\ref{tab:lasers-table}. After many photon scatters, the population will approach an equilibrium distribution $\{\bar{p}_i\}$, where $\mathrm{d}\bar{p}_i/\mathrm{d}t=0$. The equilibrium populations can be shown to have the solution
\begin{equation}\bar{p}_i = \left(\frac{f_i}{ \Gamma_i}\right)\left(\sum_j\frac{f_j}{\Gamma_j}\right)^{-1}
\label{eq:pbar-solution}
\end{equation}
by direct substitution into Eq.~\ref{eq:rate-eq}. Intuitively, the population in state $i$ is proportional to the probability $f_i$ of populating that state upon a photon scatter, as well as to the time $1/\Gamma_i$ required to pump out of the state. The sum is a normalization factor ensuring that $\sum_i\bar{p}_i=1$. The average scattering rate of molecules in the cycle is then $\Gamma_{\rm{cycle}}\equiv\sum_i \Gamma_i \bar{p}_i=\left(\sum_i  f_i/ \Gamma_i\right)^{-1}$.

We let $i=0$ represent the ground state, $\X(000)$, and $i>0$ represent repumped vibrational states. The main transition scattering rate is then $\Gamma_0\approx2$~MHz in our experiment as discussed earlier. The characteristic rate to pump into an excited state $i$ from the main transition is then $\Gamma_{\rm{in}}^{(i)}\equiv\Gamma_0 f_i$, and the characteristic rate out (for a molecule beginning in that state) is $\Gamma_{\rm{out}}^{(i)}\equiv\Gamma_i$. It is helpful to parametrize the strength of each repumping transition relative to its prominence in the optical cycle by the dimensionless ratio $\Gamma_{\rm{out}}^{(i)}/\Gamma_{\rm{in}}^{(i)}=\Gamma_i/(\Gamma_0 f_i)$. For simplicity, we consider the special case where $\Gamma_{\rm{out}}^{(i)}/\Gamma_{\rm{in}}^{(i)}\equiv \Gamma_{\rm{out}}/\Gamma_{\rm{in}}$, a fixed value for all states $i>0$.

\begin{figure}
    \centering
    \includegraphics[width=\columnwidth]{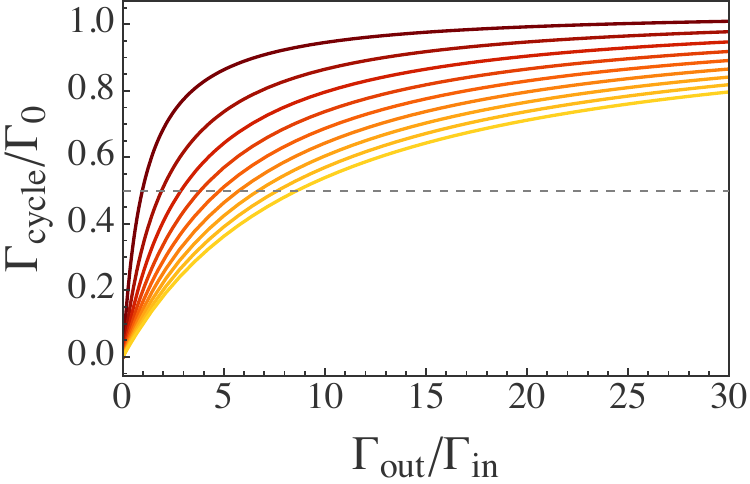}
    \caption{The scattering rate of the optical cycle, $\Gamma_{\rm{cycle}}$, relative to the main transition scattering rate, $\Gamma_0$, as a function of the characteristic repumping strength $\Gamma_{\rm{out}}/\Gamma_{\rm{in}}$ for different repumping schemes involving 2--10 lasers (top to bottom). As the number of states increases, the scattering rate requirements for each laser increase as well, matching $\Gamma_{\rm{cycle}}/\Gamma_0=0.5$ (dashed line) when $\Gamma_{\rm{out}}/\Gamma_{\rm{in}}$ equals the number of repumped states.}
    \label{fig:rates}
\end{figure}

In this special case, it follows that
\begin{equation}
    \frac{\Gamma_{\rm{cycle}}}{\Gamma_0}=\frac{\Gamma_{\rm{out}}/\Gamma_{\rm{in}}}{f_0\Gamma_{\rm{out}}/\Gamma_{\rm{in}} + n-1},
\end{equation}
where $n$ is the number of ground states included in the optical cycle. Typically, $f_0\approx1$ in an optical cycle so that if $\Gamma_{\rm{out}}/\Gamma_{\rm{in}}=1$ then $\Gamma_{\rm{cycle}}/\Gamma_0\approx1/n$, and the cycle's scattering rate is cut by the number of repumpers. Alternatively, if $\Gamma_{\rm{out}}/\Gamma_{\rm{in}}=n-1,$ the number of repumping transitions, then $\Gamma_{\rm{cycle}}/\Gamma_0\approx1/2$, and the entire cycle's scattering rate is half of the main transition's scattering rate. In Fig.~\ref{fig:rates}, we show $\Gamma_{\rm{cycle}}/\Gamma_0$ as a function of the characteristic repumping strength $\Gamma_{\rm{out}}/\Gamma_{\rm{in}}$ for various numbers of lasers included in the SrOH optical cycle. These results motivate a target scattering rate $\Gamma_{\rm{out}}/\Gamma_{\rm{in}}\gtrsim n$ from each vibrationally excited state. Based on these principles, we used $\Gamma_{\rm{out}}/\Gamma_{\rm{in}}$ as a figure of merit for each repumping transition scattering rate and aimed to ensure that $\Gamma_{\rm{out}}/\Gamma_{\rm{in}}\gg1$ for all repumps when developing our optical cycling scheme and laser system.

The model above cannot capture slowing and trapping dynamics occurring on timescales of $\sim$1--10~ms, so we expect an additional requirement that $\Gamma_i\gtrsim1$~kHz regardless of whether $\Gamma_{\rm{out}}/\Gamma_{\rm{in}}\gtrsim n$. For example, a molecule that remains in a dark state for $\sim$10~ms will likely become lost from the trap even if that state is only populated after $\sim$100~ms on average. Thus for repumping transitions occurring very late in the optical cycle, with $(\Gamma_{\rm{in}})^{-1}\gtrsim10$~ms, the required repumping rate may exceed that suggested by the simple rate equation model.

Ultimately, for our work presented in this paper, we ensured that the trapped molecule number was saturated as a function of laser power for each repumper, using scans like those shown in Figures~\ref{fig:combined-power-scans}b and \ref{fig:combined-power-scans}c. However, measuring the scattering rates was crucial to understanding the optical cycling dynamics before we could rely on direct observations of the MOT requirements.

\subsection*{Repumping Schemes for $\X(02^20)$}

When choosing a molecule for photon cycling, fewer molecular perturbations are generally preferable because perturbations can increase the number of vibrational branching channels over the expected branching caused by bond length changes. However, when repumping vibrational states, some perturbations can become useful as they can add line strength to otherwise forbidden transitions. 

As an example, CaOH has a higher number of states populated at the $10^{-5}$ level (per photon scatter) than SrOH, partly due to a Fermi resonance between the $\A(100)$ and $\tilde{A}(02^00)$ states \cite{Baum2020establishing,LasnerVBRs2022}. However, this perturbation also leads to a greater line strength for the $\X(02^20)$--$\A(100)$ transition that is used to repump $\X(02^20)$ for the CaOH MOT \cite{Baum2020establishing,vilas2021magneto}. In our early development of the SrOH MOT, we also repumped $\X(02^20)$ through $\A(100)$, but as seen in Fig.~\ref{fig:combined-power-scans}b, this transition required infeasibly high powers to drive with the desired scattering rate, and the trapped molecule number was observed to be linear in the repumping laser power. As a result, we sought an alternative repumping pathway for $\X(02^20)$. Unfortunately, no vibrational state exists within the $\B$ manifold that offers both a significant transition strength (i.e., within $\sim$1\% of a diagonal transition) and preferentially decays to another vibrational state already repumped within the optical cycle. Fortunately, the Renner-Teller interaction, which couples electronic angular momentum $\Lambda$ and vibrational angular momentum $\ell$, mixes the $\A(02^00)$ and $\tilde{A}^2\Pi_{3/2}(02^20)$ states to form the $A\mu^2\Pi_{1/2}(020)$ state (with predominantly $\A(02^00)$ composition) and $A\kappa^2\Pi_{1/2}(020)$ state (with predominantly $\tilde{A}^2\Pi_{3/2}(02^20)$ composition); see Sec. SM2. Therefore, for the results presented in this work, we repump $\X(02^20)$ through $A\mu^2\Pi_{1/2}(020)$, which decays predominantly to $\X(02^00)$. After changing to repumping through $A\mu^2\Pi_{1/2}(020)$, the trapped molecule number was fully saturated by $\sim$40~mW of $\X(02^20)$ repumping power, as seen in Fig.~\ref{fig:combined-power-scans}, whereas with the previous transition through $\A(100)$ the trapped molecule number was still linear in power even up to 1.3~W.

\end{appendix}

\bibliographystyle{apsrev4-2}
\bibliography{sroh}

\end{document}